\documentclass[
twocolumn,
superscriptaddress,
 amsmath,amssymb,
 aps,
]{revtex4-2}

\usepackage{graphicx}
\usepackage{dcolumn}
\usepackage{bm}
\usepackage[dvipsnames]{xcolor}
\usepackage{soul}
\usepackage{ulem}
\usepackage{braket}
\usepackage{multirow}
\definecolor{green}{HTML}{008000}
\definecolor{darkpurple}{HTML}{7B39A7}
\usepackage{natbib}
\usepackage{mathtools}
\begin{document}

\title{Photoinduced enhancement of chemical shift sensitivity to local vibrations}

\author{Ana Mart\'inez Guti\'errez}
\affiliation{Instituto de Ciencia de Materiales de Madrid Consejo Superior de Investigaciones Científicas (ICMM-CSIC), 28049, Madrid, Spain}

\author{Oliver Alexander}
\affiliation{Department of Physics, Blackett Laboratory, Imperial College London, SW7 2AZ London, U.K.}

\author{Pablo Est\'{e}vez Alonso}
\affiliation{Instituto de Ciencia de Materiales de Madrid Consejo Superior de Investigaciones Científicas (ICMM-CSIC), 28049, Madrid, Spain}

\author{Lorenzo Paoloni}
\affiliation{Instituto de Ciencia de Materiales de Madrid Consejo Superior de Investigaciones Científicas (ICMM-CSIC), 28049, Madrid, Spain}

\author{Terry Mullins}
\affiliation{European XFEL, Holzkoppel 4, 22869 Schenefeld, Germany}

\author{Andr\'e Al-Haddad}
\affiliation{Paul-Scherrer Institute, CH-5232 Villigen PSI, Switzerland}

\author{Thomas M. Baumann}
\affiliation{European XFEL, Holzkoppel 4, 22869 Schenefeld, Germany}

\author{Rebecca Boll}
\affiliation{European XFEL, Holzkoppel 4, 22869 Schenefeld, Germany}

\author{Christoph Bostedt}
\affiliation{LUXS Laboratory for Ultrafast X-ray Sciences, Institute of Chemical Sciences and Engineering, \'Ecole Polytechnique F\'ed\'erale de Lausanne (EPFL), CH-1015 Lausanne, Switzerland}
\affiliation{Paul-Scherrer Institute, CH-5232 Villigen PSI, Switzerland}

\author{Simon Dold}
\affiliation{European XFEL, Holzkoppel 4, 22869 Schenefeld, Germany}

\author{Alberto De Fanis}
\affiliation{European XFEL, Holzkoppel 4, 22869 Schenefeld, Germany}

\author{Gianluca Geloni}
\affiliation{European XFEL, Holzkoppel 4, 22869 Schenefeld, Germany}

\author{Markus Ilchen}
\affiliation{Deutsches Elektronen-Synchrotron DESY, Notkestr. 85, 22607 Hamburg, Germany}

\author{Iyas Ismail}
\affiliation{Laboratoire de Chimie Physique-Matière et Rayonnement, CNRS, UMR 7614, Sorbonne Université, 4 Place Jussieu, 75252 Paris}

\author{Björn Lautenschlager}
\affiliation{European XFEL, Holzkoppel 4, 22869 Schenefeld, Germany}

\author{Tommaso Mazza}
\affiliation{European XFEL, Holzkoppel 4, 22869 Schenefeld, Germany}

\author{Dooshaye Moonshiram}
\affiliation{Instituto de Ciencia de Materiales de Madrid Consejo Superior de Investigaciones Científicas (ICMM-CSIC), 28049, Madrid, Spain}

\author{Sol\`{e}ne Oberli}
\affiliation{LUXS Laboratory for Ultrafast X-ray Sciences, Institute of Chemical Sciences and Engineering, \'Ecole Polytechnique F\'ed\'erale de Lausanne (EPFL), CH-1015 Lausanne, Switzerland}
\affiliation{Laboratory of Theoretical Physical Chemistry, Institute of Chemical Sciences and Engineering, \'Ecole Polytechnique F\'ed\'erale de Lausanne (EPFL), CH-1015 Lausanne, Switzerland}

\author{Dawei Peng}
\affiliation{Laboratoire de Chimie Physique-Matière et Rayonnement, CNRS, UMR 7614, Sorbonne Université, 4 Place Jussieu, 75252 Paris}

\author{Ralph Püttner}
\affiliation{Fachbereich Physik, Freie Universität Berlin, Arnimallee 14, D-14195 Berlin, Germany}

\author{Svitozar Serkez}
\affiliation{European XFEL, Holzkoppel 4, 22869 Schenefeld, Germany}

\author{Marc Simon}
\affiliation{Laboratoire de Chimie Physique-Matière et Rayonnement, CNRS, UMR 7614, Sorbonne Université, 4 Place Jussieu, 75252 Paris}

\author{Florian Trinter}
\affiliation{Molecular Physics, Fritz-Haber-Institut der Max-Planck-Gesellschaft, Faradayweg 4-6, 14195 Berlin, Germany}

\author{Sergey Usenko}
\affiliation{European XFEL, Holzkoppel 4, 22869 Schenefeld, Germany}

\author{Michael Meyer}
\affiliation{European XFEL, Holzkoppel 4, 22869 Schenefeld, Germany}

\author{Jonathan P. Marangos}
\affiliation{Department of Physics, Blackett Laboratory, Imperial College London, SW7 2AZ London, U.K.}

\author{Jes\'us Gonz\'{a}lez-V\'{a}zquez}
\affiliation{Departamento de Qu\'{i}mica, Universidad Aut\'{o}noma de Madrid, 28049 Madrid, Spain}

\author{Daniel E. Rivas}
\thanks{Corresponding author, daniel.rivas@xfel.eu}
\affiliation{European XFEL, Holzkoppel 4, 22869 Schenefeld, Germany}

\author{Antonio Pic\'{o}n}
\thanks{Corresponding author, antonio.picon@csic.es}
\affiliation{Instituto de Ciencia de Materiales de Madrid Consejo Superior de Investigaciones Científicas (ICMM-CSIC), 28049, Madrid, Spain}

\date{\today}
\begin{abstract}
{\bf ABSTRACT:} The advent of novel free-electron laser sources enabling time-resolved x-ray photoelectron spectroscopy (tr-XPS) provides a unique opportunity to monitor local chemical environments in real time by measuring sub-eV shifts in core-electron binding energies. These shifts reflect the interplay between electronic excitation and nuclear motion, an interplay that remains largely unexplored. In our combined theoretical and experimental study of fluoropyridine (C$_5$H$_4$FN), we investigate this link by monitoring the evolving chemical environment at the N and F atomic sites as the photoexcited $S_1$ state relaxes to the ground state via a conical intersection. We find that the F site responds primarily to vibrational relaxation, showing minimal sensitivity to the electronic excited state. In contrast, excitation to $S_1$ induces a measurable energy shift at the N site and significantly enhances its sensitivity to local vibrations within the ring. This behavior arises from a photoinduced redistribution of charge, which also increases the Coulomb interaction between the 1s electron at the N atom and the atomic partial charge at an adjacent C atom. This insight opens new avenues for exploring ultrafast dynamics and conical intersection pathways in more complex systems, from photostable DNA bases to light-harvesting materials.
\end{abstract}

\maketitle


\begin{figure*}
\centering\includegraphics[scale=0.6]{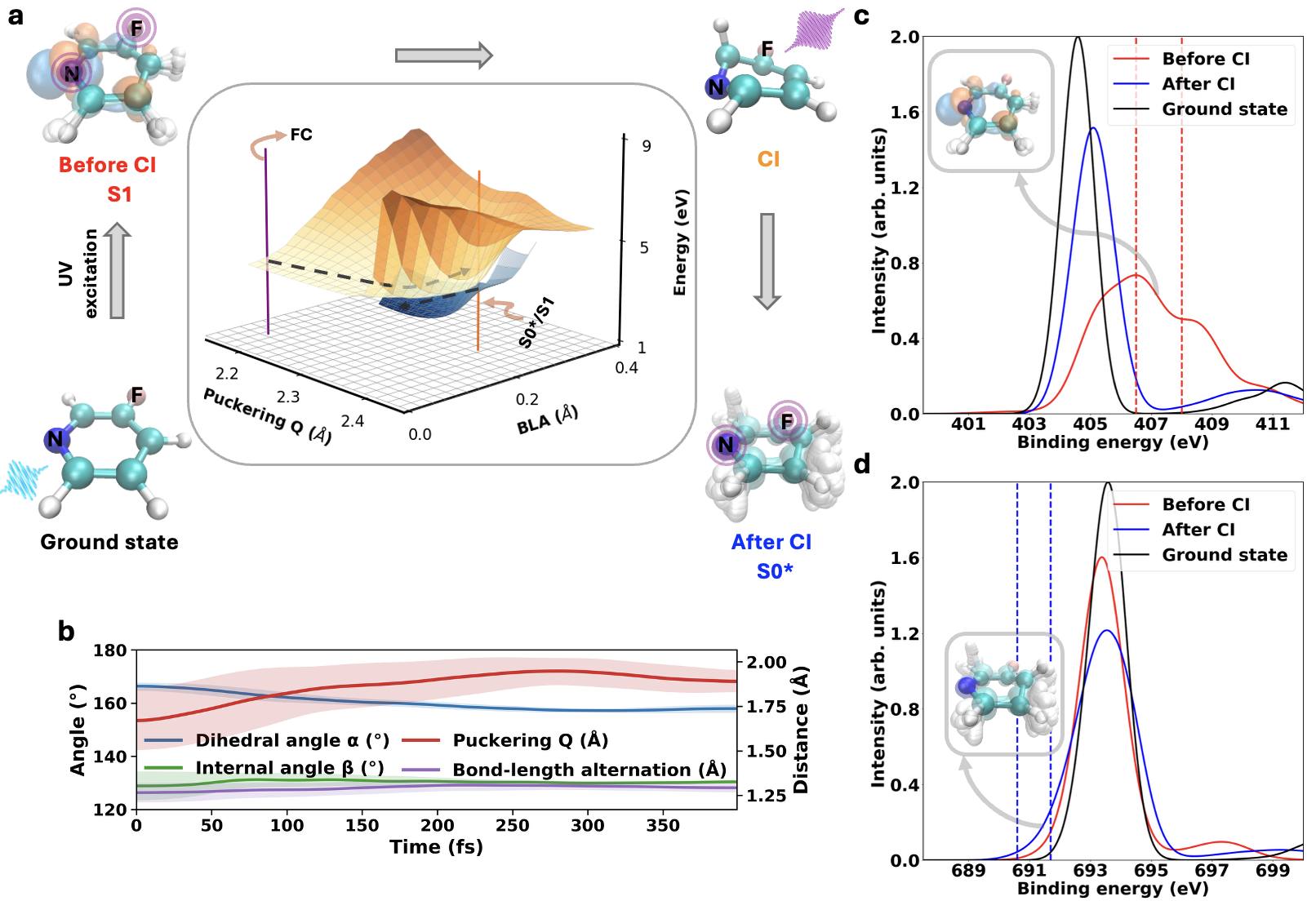}
\caption{ {\bf Chemical shifts during conical intersection passage}. (a) Schematic representation of the photoinduced vibrational relaxation of 3-fluoropyridine upon UV excitation. The UV pump pulse (cyan pulse) excites the molecule to the first excited state (before CI). The excitation is localized mainly within the ring, as evidenced by regions of increased (orange) and decreased (light blue) electron density relative to the equilibrium structure. This excitation leads to a loss of planarity, that allows a CI passage back into the ground state (after CI), through which the electronic energy converts into vibrational motion. Throughout this process, the local chemical environment at either the N atom (blue) or the F atom (red) is probed by x rays (magenta pulse), which removes electrons from their respective 1s orbitals, highlighted by circles around them. Within the schematic, the potential energy surfaces of $S_1$ and $S_0^*$ states are represented as functions of two nuclear degrees of freedom; the ring puckering amplitude (Q), which captures out-of-plane distortions of the ring, and the bond-length alternation (BLA), which reflects bond elongation at the N site. The locations of the Franck-Condon (FC) region and the conical intersection (CI) are indicated for reference. (b) Calculated time evolution of selected nuclear degrees of freedom during the relaxation process, including: the dihedral angle defined between the N–C2–C3 plane, where C3 denotes the carbon atom bonded to fluorine, and the mean molecular plane (blue line); the internal angle at the nitrogen atom formed with its two adjacent carbon atoms (green line); the ring puckering Q (red line); and the BLA (purple line). Both Q and the mean molecular plane are defined according to the original Cremer-Pople formalism \cite{cremer_general_1975}. (c) and (d) \textit{Ab-initio} calculations of the binding energies for geometries before the CI ($S_1$ electronic state, red line), after the CI ($S_0^*$ electronic state, blue line), and for the unpumped molecule (black line) at the N and F sites, respectively. Vertical dashed lines in (c) and (d) indicate the integration regions used to quantify the evolution of the $S_1$ and $S_0^*$ states, respectively.} 
\label{fig:scheme}
\end{figure*}


\section{Introduction}

When a molecule is in low-temperature equilibrium, all of its atomic constituents end up at defined average internuclear distances and charge is distributed according to the shared molecular orbitals. The chemical environment around each individual atomic site depends on the average charge around that site, which strongly varies depending on the electronegativity of the neighboring species. This was directly measured in early works on high-resolution x-ray photoelectron spectroscopy (XPS) \cite{siegbahn_electron_1982} where the shifts of the binding energy of core electrons were linked to their local chemical environment. Today, this is a widespread tool for investigating surfaces and materials with atomic sensitivity \cite{fadley_x-ray_2010,greczynski_x-ray_2020,greczynski_binding_2025}. In addition to chemical information, XPS can also reveal vibrational effects, particularly those arising from the formation of core-hole states \cite{gelius_recent_1974,hergenhahn_vibrational_2004,mendolicchio_theory_2019}. Upon ionization, the creation of a core vacancy induces significant valence electron redistribution to screen the hole, typically leading to a reduction in bond lengths and the onset of vibrational motion. Although the probing does not involve a time-resolved scheme and the initial system is in equilibrium, it remains sensitive to vibrational effects in the final states. These effects have been extensively studied and are now well understood.

Recently, with the availability of ultrashort x-ray pulses from free-electron Lasers (FELs), this technique can be extended to out-of-equilibrium systems. When a femtosecond (10$^{-15}$ s) VIS/UV pump pulse induces a reaction, the evolving chemical shifts can now be tracked in real time through the photoelectron spectra at different time delays with respect to this pump pulse, the so-called time-resolved XPS (tr-XPS). In this case, the picture may deviate from the well-established knowledge of chemical energy shifts as both excited states and nuclear motion may considerably change the charge distribution with respect to the ground state of the system.
Several excitation-induced dissociation processes have already been investigated with tr-XPS, such as Fe(CO)$_5$ \cite{leitner_time-resolved_2018}, CH$_3$I \cite{brause_time-resolved_2018}, 1-iodo-2-methylbutane \cite{allum_localized_2022}, CO \cite{al-haddad_observation_2022}, and CS$_2$ \cite{gabalski_time-resolved_2023}. Moreover, rapid electron redistribution following femtosecond excitation has been shown to manifest as measurable changes in chemical shifts  \cite{mayer_following_2022}.

If electronic excitations and nuclear distortions both influence real-time chemical shifts, an important question is how these effects evolve during dynamics governed by strong electron–nuclear coupling, for example near conical intersections (CIs). CIs are points of degeneracy between two or more potential energy surfaces, where the Born-Oppenheimer approximation breaks down and strong coupling between electronic and nuclear motion becomes critical. Such couplings play a central role in fundamental processes, including proton transfer and isomerization \cite{lennox_excited-state_2017,chen_amino_2018,gozem_theory_2017,pathak_tracking_2020}, photostability of DNA \cite{kang_intrinsic_2002}, photosynthesis \cite{cheng_dynamics_2009}, vision \cite{polli_conical_2010}, and electronic-to-vibrational energy conversion \cite{elsayed_spinorbit_1963,chergui_ultrafast_2015}. Tracking CI dynamics therefore requires a technique sensitive to both electronic excitation and nuclear motion. This poses a particular challenge for tr-XPS, as recognized in early theoretical works \cite{neville_ultrafast_2018,inhester_spectroscopic_2019}. Spectral and temporal resolution, combined with high photon energies, are required to overcome this limitation. Furthermore, probing multiple sites can provide complementary information about the dynamics \cite{CH3F}. In this context, two recent experiments in uracil \cite{facciala_unraveling_2025} and CS$_2$ \cite{thompson_shake-down_2025} reported real-time chemical shifts evolving through a conical intersection with high spectral and temporal resolution. In uracil, for example, the measurements are both sensitive to the excited state and to dynamics after CI, which returns to a vibrational hot electronic ground state. As a result, the observable captures contributions from both vibrational motion and electronic excitation, making them challenging to disentangle. Furthermore, it remains unclear whether these effects on chemical shifts act independently, i.e., simply added together, or they are intertwined. It is reasonable to expect that electronic excitation could transiently alter the chemical environment sensitivity to local nuclear distortions at specific atomic sites.

In this work, we address this question by investigating time-resolved chemical shifts along a CI passage at multiple atomic sites, combining theory and experiment. Our study focuses on a prototypical aromatic heterocycle, 3-fluoropyridine (C$_5$H$_4$FN), which is chosen as a system known for its ultrafast relaxation after electronic excitation \cite{elsayed_spinorbit_1963,yang_simultaneous_2020}. The presence of fluorine and nitrogen atoms provides distinct chemical markers located outside and within the molecular ring, respectively, enabling site-specific insight into the subsequent electronic and nuclear dynamics. A UV photon promotes the molecule to the first excited state, from which the system shows a relaxation pathway back to the ground state ($S_0$) via a CI, see Fig.~\ref{fig:scheme}(a). Through \textit{ab-initio} simulations, we show that the chemical shifts at the nitrogen site within the ring are strongly influenced by the electronic excited state and its associated nuclear dynamics, whereas the fluorine site outside the ring is primarily sensitive to molecular vibrations and largely unaffected by electronic excitation. We employ this prediction to capture the CI passage through a pump-probe experiment where we are able to independently observe this coupled electron-nuclear motion in real time and associate a decay time of approximately 1.5~ps. Finally, an analysis based on a partial charges (PC) model \cite{CH3F} reveals the interplay effects between electronic excitations and local vibrations. This shows that the chemical environment at the N atom, directly involved in the excitation, alters its sensitivity to vibrations, while more distant atoms could serve as indicators of vibrational dynamics, independently of the excited state. 

\section{Results and discussion}
A sketch of the process under investigation is shown in Fig. \ref{fig:scheme}(a). A UV (264-nm wavelength) pump pulse excites the molecule mainly to the $S_1$ excited state. In the $S_1$ state, the excitation primarily corresponds to either a $\pi\rightarrow\pi^*$ or a $n\rightarrow\pi^*$ transition, which depends on the initial molecular geometry. Here, $n$ denotes the nonbonding orbital of the nitrogen atom, while $\pi$ and $\pi^*$ represent a bonding and an antibonding delocalized orbital of the conjugated ring, respectively. After excitation, the increase in the electronic density in the $\pi^*$ orbital destabilizes the molecule, leading to its loss of planarity. When this occurs, the molecule may undergo a ring-puckering motion around the nitrogen atom \cite{cremer_general_1975}. Figure~\ref{fig:scheme}(a) shows the calculated potential energy surfaces (PESs) of the ground and first excited states. These PESs indicate that access to the conical intersection (CI) proceeds through an initial increase in the puckering amplitude (Q), defined in terms of the out-of-plane displacements of the ring atoms, followed by an increase in the bond‑length alternation (BLA), which reflects bond-length changes within the molecular ring, including those involving the nitrogen atom. This breaking of the plane symmetry causes the degeneracy of the $S_0$ and $S_1$ states at a given geometry, enabling the conical intersection passage, i.e., a radiationless transition from the $S_1$ to the $S_0$ electronic state. Based on our {\it ab-initio} simulations of the relaxation process, we analyze the key nuclear degrees of freedom. Their mean values and standard deviations over the first 350 fs are shown in Fig.~\ref{fig:scheme}(b). In particular, at early times, before CI, we observe a progressive increase in the puckering amplitude as the system evolves toward the CI. An example of a calculated trajectory during the CI is given in Section S5 of the SM. At later times, once the system returns to the $S_0$ state, the electronic energy is transformed into vibrational energy that leads to large amplitudes in the vibrational motion.

Both the evolving electronic configurations and the subsequent onset of the vibrational motion play a significant role in the localization of the transient charge, which affects the binding energies of core electrons within the system. As illustrated in Fig. \ref{fig:scheme}(a), we consider a tr-XPS scheme in which an ultrashort x-ray probe pulse (1.3-keV photon energy) ionizes the 1s electronic shells, far above the ionization threshold of N ($\sim$404 eV) and F ($\sim$694 eV) atoms. Through these photoelectrons, we obtain site-specific insight into the evolving chemical environments surrounding the F and N atoms by tracking time‑dependent shifts in their binding energies.

\begin{figure*}
\centering\includegraphics[scale=0.65 ]{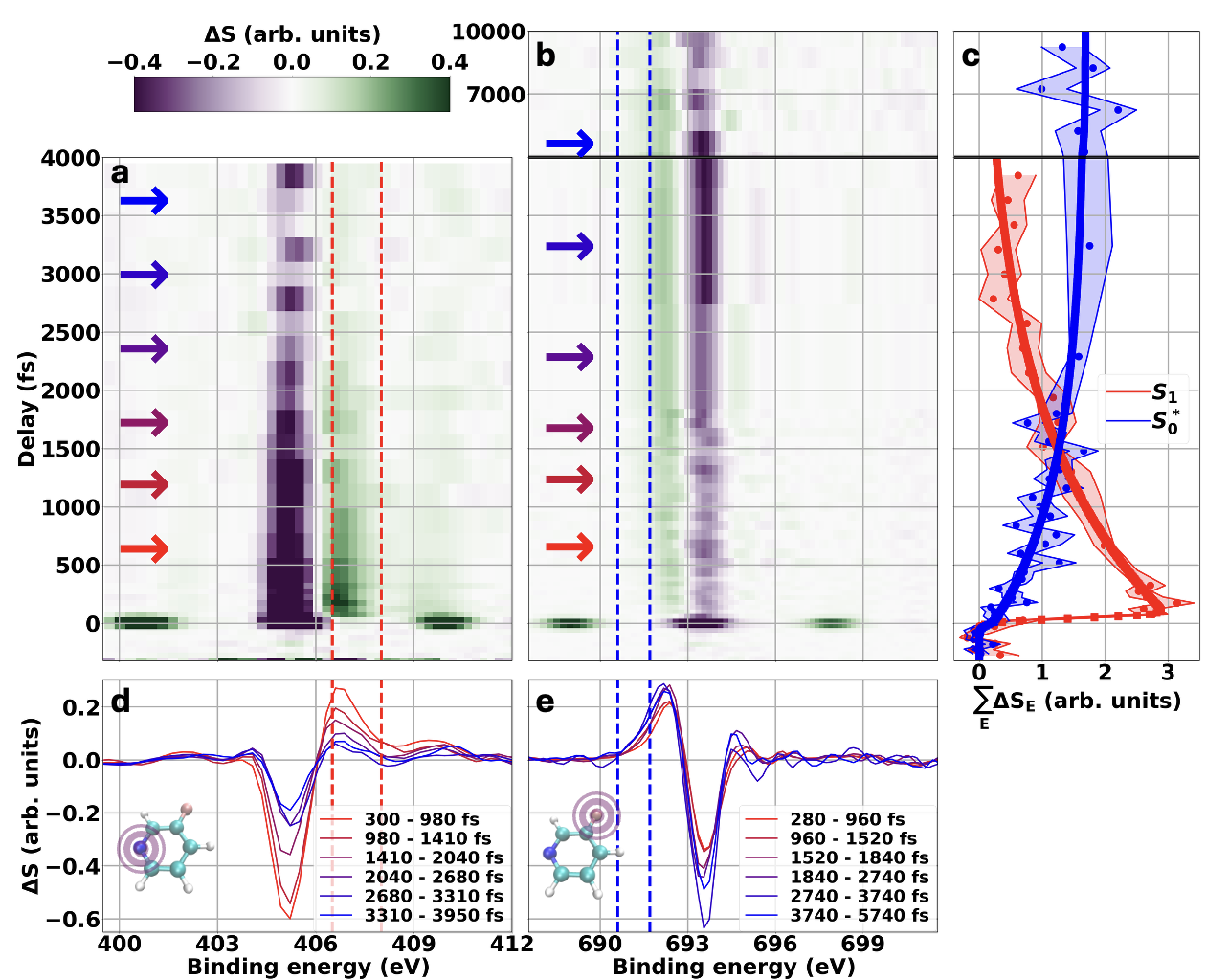}
\caption{ {\bf Tr-XPS measurements at the N and F edges.}  (a) and (b) Measured differential tr-XPS traces ($\Delta S$) up to 4 ps at the N and F sites, respectively. For the F site, an additional measurement extending to 10 ps is shown above the main trace (note the different y‑axis scale). Red and blue vertical dashed lines depict the integration regions chosen in order to characterize the evolution of $S_1$ and $S_0^*$, respectively (same regions as in Figs. \ref{fig:scheme} (a) and (b)). (c) Integrated time‑resolved difference signals within the two regions of interest indicated in panels (a) and (b). Dots represent the experimental data, while solid lines show the fit accounting for the photoexcited dynamics, see main text. Shaded areas denote the standard deviation at each time delay. (d) and (e) Difference XPS lineouts over selected time-delay ranges at the N and F sites, respectively. Colored arrows in panels (a) and (b) mark the corresponding time‑delay intervals.}
\label{fig:expdata}
\end{figure*}

We first focus on the N site. Here, we probe within the ring, at the site of the delocalized $\pi$ orbitals, which plays a fundamental role in the ring puckering towards the CI. We perform advanced \textit{ab-initio} calculations of the excited-state dynamics and separately calculate the corresponding binding energy of this atom at each time step. The static XPS calculations are benchmarked with synchrotron measurements performed at the GALAXIES beamline of Synchrotron SOLEIL \cite{ceolin_hard_2013}, see details in Section  S3-C of the Supplementary Material (SM). Our theory is based on a semiclassical approach that treats the nuclear motion classically, while the electronic structure is treated at the quantum level (see section D of Methods for more details). We run several semiclassical trajectories starting in the $S_1$ state, modeling the UV excitation, and simulate the dynamics that accounts for non-adiabatic couplings, up to 1000 fs. At the end of the simulation, we obtain a set of trajectories, some of which have undergone the CI passage. For each trajectory, we group all instantaneous geometries as `before CI' and `after CI', separated by the time-step when the CI passage occurred. The respective calculated average binding energies at the N site for these groups, together with the binding energy of the ground-state molecule, are shown in Fig. \ref{fig:scheme}(c). 

On the one hand, for the geometries corresponding to `before CI', we observe a strong depletion of the main peak around 404.5 eV and an increase of signal around 406.5-408 eV with respect to the ground state [see Fig. \ref{fig:scheme}(c)]. The ground‑state broadening reflects the nuclear distribution of the equilibrium wavefunction. When compared with this ground‑state width (1.25 eV FWHM), the widening of binding energy shifts before reaching the CI is particularly striking. The region around 409 and 412 eV is primarily attributed to photoelectron satellites, arising from core ionization accompanied by valence excitations. Throughout the excited-state $S_1$ dynamics, the $S_1$ and $S_2$ states remain energetically close, and non-adiabatic coupling mixes them during propagation (see an example in Section S6-A of the SM). These states exhibit $\pi\pi^*$ and $n\pi^*$ electronic character, which becomes strongly mixed during evolution. Vibrations `before CI' are less pronounced than `after CI', see illustration Fig.~\ref{fig:scheme}(a); therefore, the observed energy-shift widening is attributed to an intertwined effect of the electronic configuration change and the vibrational motion. Consequently, the binding energies in the 406.5-408.0 eV range [indicated by the red vertical dashed lines in Fig. \ref{fig:scheme}(c)] can be directly associated with the $S_1$ dynamics. On the other hand, for the geometries labeled `after CI', although the system ends up at the ground electronic level ($S_0$), we predict a depletion of the main peak, a slight shift towards higher binding energies, and a broadening from 1.25 eV to 1.5 eV relative to the ground state. We attribute this broadening to vibrations arising from relaxation to the vibrationally hot $S_0^*$ state. However, disentangling this broadening from the large contribution of $S_1$ is challenging, making direct measurement of the $S_0^*$ population at the N site difficult. As we shall show in the next paragraph, this is not the case at the F site.

Similarly to the N site, we calculate the F binding energy for the simulated trajectories and then group the geometries into `before CI' and `after CI' [see Fig. \ref{fig:scheme}(d)]. Now, in the `before CI' trace, we observe a slight broadening and energy shift towards lower binding energies. Additionally, there is a slight increase at 697 eV associated with a chemical shift of the satellite. However, in the `after CI' case, the broadening is stronger than in the `before CI' case. As this broadening is associated with molecular vibrations, the spectral region 690.6-691.7 eV [as indicated by the blue vertical dashed lines in Fig. \ref{fig:scheme}(d)] will be sensitive to observe the onset of energy conversion to nuclear degrees of freedom. 

From this theoretical modeling, we rationalize the distinct effects of the CI passage at two atomic sites. Although the calculated XPS signals strongly overlap, we identify that an increased signal toward higher binding energies near 406.5-408.0 eV at the N site can be mainly associated with the excited-state population $S_1$ and an increased signal near 690.6-691.7 eV at the F site is associated with the vibrationally excited ground-state population $S_0^*$. At both edges, we also observe significant changes in satellite features at larger binding energies (above 409 eV for N and above 695 eV for F). While these satellite states are not explored further here, we predict that such signals could be used to disentangle dynamics in more complex systems; see, for example, \cite{thompson_shake-down_2025}. We now turn to see how this concept is applied to an experiment. 

We performed the experiment at the Small Quantum System (SQS) instrument of the European XFEL. UV laser pulses of 264 nm central wavelength were used to excite the molecules, and monochromatized x-ray pulses of 1.3 keV photon energy and a bandwidth of 0.67 eV were used as a probe (see section A of Methods for additional details). The x-ray pulses ionize the 1s orbitals of both nitrogen and fluorine (binding energies of around 405 and 693 eV, respectively) allowing to investigate both atomic sites without varying the x-ray parameters. Both the pump and probe lasers are polarized along the horizontal axis. The photoelectron spectrum along the polarization direction is measured with an electron time-of-flight spectrometer in the Atomic-like Quantum Systems (AQS) end-station \cite{de_fanis_high-resolution_2022}.  A retardation potential of 850 V and 570 V was applied to the spectrometer, decelerating the electrons to obtain a high-energy resolution at either the nitrogen or the fluorine edge, respectively. By varying the time delay between pump and probe pulses, we obtain the tr-XPS traces.

For better interpretation of the evolving transient XPS signals with respect to the ground state, we show the normalized difference, $\Delta S$, between the photoelectron spectra of the pumped and unpumped molecules (see section B of Methods for additional details). The obtained $\Delta S$ traces for the N and F sites are shown in Figs. \ref{fig:expdata} (a) and (b), respectively. 

We begin by discussing the measurement at the N site. When pump and probe pulses overlap in time (delay = 0 fs), we observe a strong depletion centered around 405 eV accompanied by an increase in signal at $\pm 4.7$ eV from the main photoline. This is due to the dressing effect of the optical laser, a nonlinear process in which the outgoing electron gains or loses a single quanta of energy from the laser field. From the effective duration of the dressing effect, we obtain an experimental temporal resolution of 64.8$\pm$1.3~fs, which is mainly determined by the duration of the UV pulse (see Fig. S4-b of the SM). For positive time delays (pump before probe) we recognize a strong shift toward larger binding energies, seen as a depletion between 405 and 406 eV and an increase of signal between 406.5 and 408 eV. To better visualize these features, we perform lineouts of the evolving $\Delta S$ at selected time delays [see Fig. \ref{fig:expdata}(d)]. We observe that both the positive and negative signals decrease in amplitude over time. Based on the results of our theoretical model, we understand that this behavior corresponds to evolving population of the $S_1$ state, as described in Fig. \ref{fig:scheme}(b). There we predict that the region between 406.5 and 408 eV [between the red dashed lines shown in Fig. \ref{fig:expdata}(a)] would be a good representation of the $S_1$ population, so we integrate $\Delta S$ in that region for each time step. The results [red dots in Fig. \ref{fig:expdata}(c)] show the decreasing signal in time, representative of the $S_1$ population. However, from the results presented in Fig. \ref{fig:scheme}(c), we can see that some small contribution from the $S_0^*$ state could be present. We therefore perform a fit that considers the decreasing $S_1$ population and increasing $S_0^*$ population, assigning initial weights to each state (see section D of Methods for details). Based on this fit, we obtain a time constant of $1530 \pm 390 $ fs for the CI passage, consistent with values reported previously for pyridine and related heteroaromatic systems \cite{yang_simultaneous_2020,helle_studies_2022,feng_excitation_2023}. We emphasize that this approach provides unprecedented spatial and temporal resolution in resolving the dynamics through the conical intersection. Furthermore, from the fitted relative population weights, we find that the $S_0^*$ state only contributes $2\%$ to the difference in the chosen spectral region, which confirms our initial prediction based on the theoretical \textit{ab-initio} calculations. 

Now turning to the F site [see Fig. \ref{fig:expdata}(b)], we observe a chemical shift in the opposite direction for positive time delays, as predicted by our theoretical modeling. The differential signal is dominated by a negative signal between 693 and 694 eV and positive signals between 690 and 693 eV. We perform lineouts of selected time windows for better visualization of the evolving trace [see Fig.~\ref{fig:expdata}(e)], where we note that these signals increase in amplitude over time. Based on the theoretical model shown in Fig. \ref{fig:scheme}(d), we associate these traces with the build-up of $S_0^*$ population in time. At each time step, we integrate $\Delta S$ over the 690.6-691.7eV range, indicated by the blue dashed lines shown in  Fig. \ref{fig:expdata}(b) [also region highlighted in blue in Fig. \ref{fig:scheme}(d)]. The results [blue dots in Fig. \ref{fig:expdata}(e)] now show a signal that increases over time. Again considering possible contributions from both increasing $S_0^*$ population and decreasing $S_1$ population, we perform the same fitting procedure used for the N site. From this fit, we obtain a time constant of $1210 \pm 280 $ fs, consistent with the result obtained from the N site. Hence, most of the excited population ultimately returns to $S_0^*$. The relative contribution of $S_1$ to this signal is $23\%$. We note that while the integration windows could, in principle, be broadened, we deliberately restrict them to regions that remain predominantly sensitive to the `after-CI' population without compromising the signal statistics. At higher binding energies, the `before‑CI' state contributes more strongly, as illustrated in Fig. \ref{fig:scheme}(d).

We demonstrate that tr-XPS can successfully capture features arising from $S_1$ and $S_0^*$ dynamics during passage through a CI. Our results highlight the advantage of high spectral resolution for resolving time-dependent chemical shifts and isolating the relevant dynamics. As shown, accurate modeling of these dynamics is essential to guide and interpret the measurements. However, \textit{ab-initio} calculations of binding energies are computationally demanding. In the next section, we will show that further insight on the dynamics can be obtained using a partial charges model \cite{CH3F}, a significantly less computationally intensive approach. So far, we attribute the broad spectral features at the N site to the $S_1$ dynamics, which \textit{ab-initio} calculations suggest arise from an intertwined effect of electronic excitation and vibrational motion. In contrast, this effect is absent at the F site, where time-dependent features appear to evolve primarily due to the excitation of a larger number of vibrational levels, with minimal influence from the change of electronic state when the system goes through the CI. In the following section, we further explore this phenomenon by using the methodology detailed in Ref. \cite{CH3F}.

\begin{figure*}
\centering\includegraphics[scale=0.58]{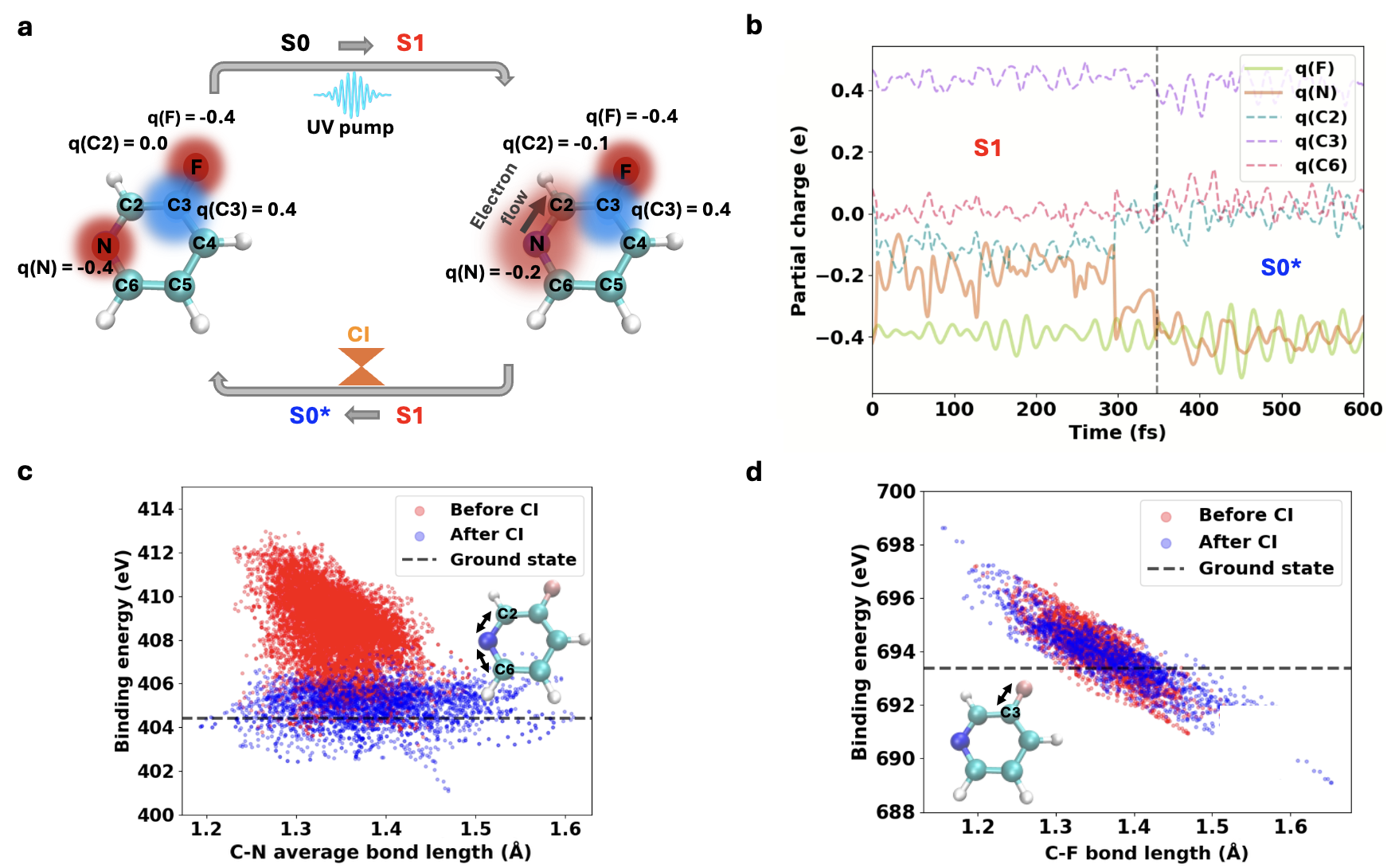}
\caption{{\bf Photoinduced enhancement of chemical shift sensitivity to local vibrations.} (a) Illustration of the atomic charge distribution in 3-fluoropyridine during the photoexcited dynamics, where blue denotes positive charge (lower electron density) and red denotes negative charge (higher electron density). Upon excitation from the $S_0$ to the $S_1$ state, electron density flows from the nitrogen atom toward the adjacent carbon atom (C2). After passage through the conical intersection, this charge redistribution is reversed, restoring the ground‑state pattern. (b) Time evolution of the partial atomic charges for the most relevant atoms contributing to the chemical shifts at the F and N sites, shown for a representative semiclassical trajectory. The passage through the conical intersection occurs at approximately 350 fs, as indicated by the vertical dashed black line. A simplified PC model that considers only neighboring atoms is used to calculate binding energies for each trajectory. (c) and (d) Binding energies as a function of the relevant bond‑length coordinates for all computed trajectories that undergo relaxation, at the N site with neighboring carbons C2 and C6 (c), and at the F site with carbon C3 (d). Geometries before and after CI are distinguished. These plots show that chemical shifts at the F site do not depend on the electronic state, whereas at the N site their sensitivity to nuclear distortions strongly depends on the electronic excitation.
}
\label{fig:partial_charges}
\end{figure*}

Chemical shifts, in equilibrium systems, are effectively captured by a partial charges model \cite{gelius_molecular_1970,gelius_binding_1974}. A PC model estimates chemical energy shifts through an electrostatic potential, whose form depends mainly on the effective partial charges at each atomic site and the distances between the ionized site and its neighboring atoms (see Section E of Methods for more details). This model is well-known for its application to static XPS \cite{gelius_molecular_1970}, but in a previous work we showed that a PC model is also able to capture dynamical chemical shifts, see Ref. \cite{CH3F}. For 3-fluoropyridine, there is also a remarkably good agreement between \textit{ab-initio} binding energies and those calculated from a PC model, see more details in Section S4 of the SM. Using the PC model, we identify that the most influential parameters governing the chemical shifts are the terms that depend on the ionized site and its nearest neighbors. This indicates that the main trends in dynamical chemical shifts can be captured by accounting solely for vibrations of neighboring bonds and their associated partial charges. 

We observe a pronounced charge redistribution immediately after photoexcitation around the N atom, whereas no significant change occurs near the F atom, as depicted in Fig. \ref{fig:partial_charges}(a). The loss of electron density at the N site, accompanied by an increase on an adjacent C atom, alters the sensitivity of the local chemical environment, where Coulomb interactions between neighboring atomic charges also become more relevant. Figure \ref{fig:partial_charges}(b) shows the evolution of partial charges at the F, N, and their adjacent C atoms for one selected trajectory. At the F atom and its neighboring C atom (C3), the partial charges oscillate around –0.4{\it e} and +0.4{\it e}, respectively, and remain on average unchanged after the CI passage at 350 fs. The only noticeable effect is an increase in the oscillation amplitude, driven by bond‑length variations associated with the higher vibrational energy acquired upon return to the $S_0$ state. In contrast, the N atom exhibits a different behavior: after photoexcitation at 0 fs, its atomic charge increases by +0.2{\it e} in $S_1$ from its initial value -0.4{\it e} in $S_0$, losing this excess of positive charge after the CI passage (note that $S_0^*$ retains the same average partial charge distribution as $S_0$). Among the adjacent C atoms, one remains nearly unaffected (C6), while the other (C2) receives approximately –0.1{\it e} immediately after photoexcitation, which returns to zero following the CI passage. This transient charge flow between the N and C6 atoms enables the Coulomb interaction between them, thereby enhancing the sensitivity of the N-site chemical environment to internuclear distances exclusively in the $S_1$ state. Notably, variations in the partial charges at the nitrogen and adjacent carbon atoms are already evident after 300 fs, as the system evolves toward the CI and the molecular geometry undergoes significant puckering distortion.

To highlight the sensitivity of chemical shifts to vibrational motion, we analyze the variation of the binding energy across all semiclassical trajectories, separating between before and after CI, as a function of the neighboring bond lengths of the ionized site, see Figs. \ref{fig:partial_charges}(c) and (d) for the N site and F site, respectively. 
At the N site, there is a pronounced difference between molecules that have crossed the CI and those that have not. Before the CI passage, changes in the C-N bond length lead to significant variations in chemical shifts. After the CI passage, this sensitivity becomes much weaker. Although vibrations increase after passing the CI, the PC model indicates that they no longer drive substantial changes in local charges, and the Coulomb interaction term for the neighboring atom is practically zero, unlike in the state $S_1$. This change in sensitivity is linked to the $S_1$ excitation, which involves molecular orbitals within the ring where the N atom resides. This also explains the broader signal observed in Fig. \ref{fig:scheme}(c) for the `before CI' case compared to the ground state or the `after CI' signal.
At the F site, there is a direct correlation between the C-F bond length and the chemical shift. Here, vibrations consistently influence local charge and the Coulomb interaction term in the PC model, regardless of the electronic state. Notably, the local charge at the F site varies strongly with bond length, but the chemical shift is unaffected by the excitation. This confirms the expectation that F serves as a reliable marker outside the ring for tracking the return to the vibrationally excited state $S_0^*$.

\section{Conclusions}
In summary, we have investigated the ultrafast passage through a conical intersection in a prototypical heterocyclic compound, 3-fluoropyridine, through multi-site tr-XPS. Upon UV excitation, the molecule is promoted to the first excited state $S_1$, and our experimental data reveals that the excited population returns to a vibrationally hot ground state $S_0^*$ with a decay time of approximately 1.5 ps. Our theoretical modeling of these results reveal that the time-dependent chemical shifts at the F site are strongly influenced by C-F bond length variations and its sensitivity is largely unaffected by the electronic state. In contrast, shifts at the N site during $S_1$ evolution stem from an interplay between electronic state and vibrational motion that redistributes electron density. This leads to larger chemical shifts before the conical intersection than after, even though vibrational amplitudes are significantly greater following the CI passage. This effect is attributed to a charge flow between the N site and an adjacent carbon following light excitation, which involves molecular orbitals within the ring with $n$ and $\pi$ character. This charge redistribution increases the sensitivity to C-N bond length variations, in which the Coulomb interactions between neighbor atomic charges play an important role. By demonstrating sensitivity to vibrational motion at electronically excited active sites, this study opens new avenues for probing coupled nuclear-electronic dynamics in increasingly complex environments.



%

\section{Methods}

\subsection{Experimental setup}

The XFEL was tuned such that the FEL gain process is not saturated, aiming to produce pulses below 10 fs. An energy of approximately 2 mJ is delivered before the monochromator. The bandwidth is later reduced to 0.67 $\pm$ 0.1 eV with the dedicated grating-based soft x-ray monochromator located in the SASE3 beamline of the European XFEL \cite{mono}, leading to an energy of approximately 1.5 µJ being delivered to the SQS beamline. At the interaction point, a focal spot of a few micrometers is expected, much smaller than the UV focal spot of 96 $\pm$ 10 µm FWHM. The 264-nm UV pulses were generated by first frequency doubling the fundamental wavelength of the dedicated optical laser, and then generating the third harmonic via sum-frequency generation. A set of dispersive mirrors providing -145 fs$^2$ per bounce (DM100, Ultrafast Innovations) were employed to compensate for the chirp added in the UV-generation beam path. A total energy of 3.2 $\pm$ 0.1 µJ is delivered to the interaction region. 

From the delay dependence of sidebands in the photoelectron spectra, which are proportional to the overlap between the UV-laser and x-ray pulses, we can determine that the overall temporal resolution is approximately 65 fs. This is mainly due to the UV pulse duration, but also has contributions from the XFEL pulse duration and the residual time jitter after bunch-arrival corrections (see section S1-C of the SM for further details).

The fluoropyridine molecules, liquid at room temperature, are directly delivered to the chamber via a needle, whose exit is positioned a few millimeters from the interaction point. The low vapor pressure of the sample allows it to be degassed by directly pumping on it at room temperature. A needle valve is used to control the sample delivery, maintaining a background pressure of $2\times 10^{-6}$ mbar in the interaction chamber.

\subsection{Differential photoelectron spectra}

The XFEL was operated such that trains of electron bunches were produced at a repetition rate of 10 Hz. Each train contained 134 lasing bunches, producing x-ray pulses with an intra-bunch repetition rate of 376 kHz. This was chosen as double the repetition rate of the UV laser in order to have an equal number of pump-probe versus `un-pumped' measurements. Doing this in an interleaved way allows any systematic drifts that might occur along the XFEL pulse train to be minimized. 

For each x-ray pulse that coincides with a UV pulse (every other event), the delay between the x-ray and UV pulses is determined by combining measurements on a bunch arrival monitor, the delay stage, and referencing to the sidebands produced in the photoelectron spectra at temporal overlap (see section S1-B of SM for additional information). 

We define delay bins $i$ with $\tau_i$ delay at the center of the bin, and sum the photoelectron spectra of all x-ray-UV pulse pairs to retrieve total photoelectron kinetic energy (K.E, directly related to the binding energy) spectra for each delay bin, $S_{\mathrm{pump}}(E_{\mathrm{K.E.}}; \tau_i)$. The same is done for x-ray pulses that do not coincide with UV pulses, using the delay of the pulse pair immediately preceding each pulse, to find the analogous total photoelectron kinetic energy spectra, $S_{\mathrm{0}}(E_{\mathrm{K.E.}}; \tau_i)$.

We obtain the differential tr-XPS trace at $\tau_i$ delay by normalizing the pumped and unpumped total photoelectron spectra to the total pulse energy contributing to them, $I_{\mathrm{pump/0}}$, and taking their difference:

\begin{equation}
    \label{eqn:exp-diff}
    \Delta S(E_{\mathrm{K.E.}}; \tau_i) = \frac{S_{\mathrm{pump}}(E_{\mathrm{K.E.}}; \tau_i)}{I_{\mathrm{pump}}(\tau_i)} - \frac{S_{\mathrm{0}}(E_{\mathrm{K.E.}}; \tau_i)}{I_{\mathrm{0}}(\tau_i)}.
\end{equation}
More details can be found in Section S1 of the SM.


\subsection{Fitting of differential tr-XPS traces}

To quantify the transitions between states following excitation, we integrate regions of the XPS traces and fit to their time dependence. 
Changes in the spectra are caused by time dependence of the population in the ground state, before the CI and after the CI. 

To understand the choice of function, we first consider an infinitely short pump pulse.
In regions of binding energy where there is a difference between the ground state and before-CI photoelectron spectra, we expect a step change in the spectrum at temporal overlap, followed by an exponential decay of this change as the molecules pass through the CI.
In regions of the binding energy where there is a difference between the ground state and after-CI spectra, we expect a step change in the gradient at temporal overlap followed by the onset of signal of the form $1-e^{-\tau}$ as the population passes through the conical intersection.
In general, each region of the spectra can have contributions from both $S_1$ (before) and $S_0^*$ (after) and is described by
{\small
\begin{equation}
    f(\tau) = \!
    \begin{dcases}
    0 & \!\! \tau < \tau_0 \\
     A_{\mathrm{before}} e^{-(\tau-\tau_0)/\tau_{\mathrm{CI}}} + A_{\mathrm{after}}\left(1 - e^{-(\tau-\tau_0)/\tau_{\mathrm{CI}}}\right) & \!\! \tau \geq \tau_0
    \end{dcases},
\end{equation}
}
where $\tau_{\mathrm{CI}}$ is the rate of population transfer through the conical intersection, $\tau_0$ is the offset in the change from temporal overlap, and $A_{\mathrm{before}}$ and $A_{\mathrm{after}}$ are the differences in amplitude in that region of the XPS from the ground state for states before and after the conical intersection, respectively.

To account for temporal broadening due to the finite time resolution of the experiment, we fit to the data a convolution of this function with a Gaussian function with width matching the resolution, i.e., 64.8 fs FWHM:
\begin{equation}
    g(\tau) = \int \frac{1}{38.9 \sqrt{2\pi}}\exp \left(-\frac{\tau'[\mathrm{fs}]^2 }{38.9^2}\right) f(\tau - \tau') d\tau'.
\end{equation}
$A_{\mathrm{before}}$, $A_{\mathrm{after}}$, $\tau_{\mathrm{CI}}$, and $\tau_0$ are free parameters.

\subsection{\textit{Ab-initio} model for transient chemical shifts}

We developed an \textit{ab-initio} model to describe the dynamics and calculate the XPS spectra based on a semiclassical surface-hopping model \cite{tully_molecular_1990}. In our model, the electron motion is described at the quantum level, while the nuclear dynamics is treated classically using a swarm of trajectories to mimic the nuclear wavepacket motion, see more details in Section S2-C of the SM. We use a version implemented in SHARC \cite{richter_sharc_2011} to perform the semiclassical nuclear dynamics.

The electronic structure calculations for the ground and low‑lying excited states were performed using a multireference approach, specifically the complete-active-space self-consistent field method with an active space of 8 electrons and 7 orbitals (CASSCF(8,7)). Dynamical electron correlation effects were accounted for by applying energy corrections using the Complete Active Space Second-Order Perturbation Theory (CASPT2) method \cite{shiozaki_communication_2011}. The electronic calculations during the dynamics are performed at the CASPT2 level of theory with the cc-PVTZ Dunning basis set using the BAGEL software \cite{shiozaki_bagel_2018}.

The absorption of the UV photon is modeled by the excitation of the vibrational ground state into the $S_1$ and $S_2$ electronic excited states. The excited states are populated by considering first-order perturbation theory, in which the $S_1$ is mainly populated (see more details in Section S5 of the SM). The trajectories start then in an excited state and are propagated using a surface-hopping algorithm, which incorporates nonadiabatic couplings in order to describe the conical intersection. A total of 68 trajectories with different initial conditions are propagated to ensure good statistics and capture the main dynamics for approximately 1 ps. Within 800 fs, around $20\%$  of the trajectories relaxed to the ground state (see Fig. S15-b in Section S5 of the SM). The initial conditions, comprising coordinates and velocities for each trajectory, are sampled from an harmonic Wigner distribution in the ground electronic state, which reproduces quantum distributions in both position and momentum space.

The electronic calculations of core-hole or core-ionized states, in which the 1s orbital is included in the active space, are performed to determine the binding energies and Dyson intensities \cite{ortiz_dyson-orbital_2020}. The 1s orbital is incorporated in the first restricted active space (RAS1) via a Restricted Active Space Self-Consisted Field (RASSCF) calculation, carried out using the OpenMolcas software \cite{fdez_galvan_openmolcas_2019,aquilante_modern_2020}. Energies are corrected at the CASPT2 level of theory. We performed a state average over 30 states and 50 states for the F edge and N edge respectively, in order to achieve maximum accuracy in both the main and satellites signals. We evaluated the accuracy of this level of theory by calculating the static XPS of 3-fluoropyridine and comparing it with experimental data acquired at the SOLEIL synchrotron at the GALAXIES beamline \cite{ceolin_hard_2013}. This enables us to establish the use of a cc-PVTZ Dunning basis set with an excellent agreement with the static XPS experiment (see the calculated static XPS in Fig. S10 of the SM).

For each trajectory obtained during the nuclear propagation, an incoherent sum of the ionization intensities is computed over all geometries to construct the transient XPS signal. The ionization yield is evaluated using Dyson intensities, defined as the norm of the Dyson orbitals between the lowest-energy states and the core-ionized states. These orbitals are calculated at the CASPT2 level using the OpenMolcas software package. In all computed \textit{ab-initio} binding energies we apply an absolute energy shift of 1.1 eV and 0.7 eV at the F and the N edges, respectively, to compare with the experimental data. In those trajectories propagated in the ground state to compute Fig. \ref{fig:scheme} we apply an additional relative shift of –0.2 eV at the F edge.

Figures of the molecular structure, electron density, and molecular orbitals were generated using the Visual Molecular Dynamics (VMD) software package \cite{humphrey_vmd_1996}.

\subsection{Partial charge model}

We analyze the transient charge distribution in the 3-fluoropyridine molecule. Out-of-equilibrium chemical shifts can be accurately described using a partial charge model expressed as in Ref. \cite{CH3F}:~$\Delta E_{A_{i}}=k_{A_{i}}\cdot q_{A}+ \sum_{B\neq A}\frac{q_{B}}{R_{AB}} +l_{A_{i}}$ where $A$ denotes the core-ionized atom, $B$ represents any other atom in the molecular system, and $R_{AB}$ is the distance between atoms $A$ and $B$. The term $q_{A}$ is the partial charge on atom $A$, and $i$ refers to the core-ionized orbital of  atom $A$. The quantity $\Delta E_{A_{i}}$ corresponds to the chemical shift of the core orbital $i$ relative to a reference level. The constant $k_{A_{i}}$ represents the average Coulomb repulsion between the core electron $i$ of atom $A$ and its valence electrons, while $l_{A_{i}}$ introduces an absolute energy offset. Mulliken or L\"owdin partial charges provide a reasonable description at the F site but fail at the N site due to basis set delocalization within the ring \cite{saha_are_2009}. To address this limitation, we perform a Natural Population Analysis (NPA) using the JANPA software \cite{Nikolaienko_janpa_2014} on the natural orbitals from the \textit{ab-initio} dynamical simulations at the level of CASSCF(8,7) theory, which allows us to extract reliable atomic partial charges in both sites. We use $k=32.5$ eV/C at the F site \cite{davis_x-ray_1972} and $k=27.0$ eV/C at the N site \cite{brown_n1s_1980}. These values remain essentially constant during the dynamics because the core orbitals are highly localized.

\section*{Acknowledgments}
The authors gratefully acknowledge the support teams at European XFEL and DESY for their assistance in conducting this experiment under proposal reference 2862. A.M.G, P.E.A, L.P., D.M., and A.P. acknowledge the Spanish Ministry of Science, Innovation and Universities \& the State Research Agency through grants refs. PID2021-126560NB-I00, PID2024-157663NB-I00, CNS2022-135803, PID2022-143013OB-100, and CNS2023-145046 (MCIU/AEI/FEDER, UE), and the "Severo Ochoa" Programme for Units of Excellence in R\&D (CEX2024-001445-S), and computer resources and assistance provided by the Red Espa\~nola de Supercomputaci\'on (RES) under projects refs. FI-2024-3-0011, FI-2024-2-0034, FI-2023-2-0012, FI-2023-1-0035, FI-2022-3-0022, and FI-2022-1-0031. This publication is based upon work from COST Action NEXT, CA22148 supported by COST (European Cooperation in Science and Technology). S.O. ackowledges support from the Swiss National Science Foundation, National Center of Competence in Research - Molecular Ultrafast Science and Technology NCCR - MUST. T.M. and M.M. acknowledge support from the DFG, the German Research Foundation (Deutsche Forschungsgemeinschaft, Project 170620586, SFB-925). M.I. acknowledges support by the Bundesministerium für Bildung und Forschung (BMBF) Grant No. 13K22CHA, the Cluster of Excellence “Advanced Imaging of Matter” of the Deutsche Forschungsgemeinschaft (DFG)-EXC 2056—Project ID No. 390715994, and (DFG)-Project No. 328961117-SFB 1319 ELCH (“Extreme light for sensing and driving molecular chirality”). O.A. and J.P.M. wish to acknowledge funding from UK EPSRC grant numbers EP/X026094/1 and EP/V026690/1. F.T. acknowledges funding by the Deutsche Forschungsgemeinschaft (DFG, German Research Foundation) -- Project 509471550, Emmy Noether Programme.

\section*{Supporting information}

Additional experimental details, calculations, analysis, and methods, including additional Figures S1-S27.
A movie of the representative trajectory is available in the supporting data for better visualization of the dynamics.

\section*{Data availability}
Data underlying the results presented in this work will be available in red {https://doi.org/10.20350/digitalCSIC/18337} and {https://doi.org/10.22003/XFEL.EU-DATA-002862-00.}

\clearpage

\bibliography{references}

\end{document}